\documentclass{article}
\usepackage{arxiv}

\usepackage{mathrsfs}

\usepackage{caption}
\usepackage{subcaption}
\usepackage{listings}
\usepackage{paralist}
\usepackage{framed}
\usepackage[boxed]{algorithm2e}
\usepackage{amsmath}
\usepackage{color}

\usepackage{url}
\usepackage{todonotes}

\usepackage{amsmath}

\title{A Novel IaaS Tax Model as Leverage Towards Green Cloud Computing}

\author{Benedikt Pittl\\
	Faculty of Computer Science\\
	University of Vienna\\
	A-1090 Vienna, Austria\\
	\texttt{benedikt.pittl@univie.ac.at} \\
	\And
	Werner Mach\\
	Faculty of Computer Science\\
	University of Vienna\\
	A-1090 Vienna, Austria\\
	\texttt{werner.mach@univie.ac.at} \\
    \AND
    Erich Schikuta\\
	Faculty of Computer Science\\
	University of Vienna\\
	A-1090 Vienna, Austria\\
	\texttt{erich.schikuta@univie.ac.at} \\
}

\begin{document}

\date{}

\maketitle

\begin{abstract}
The cloud computing technology uses datacenters, which require energy. Recent trends show that the required energy for these datacenters will rise over time, or at least remain constant. Hence, the scientific community developed different algorithms, architectures, and approaches for improving the energy efficiency of cloud datacenters, which are summarized under the umbrella term Green Cloud computing. 
In this paper, we use an economic approach - taxes -  for reducing the energy consumption of datacenters. We developed a tax model called~\emph{GreenCloud tax} which penalizes energy-inefficient datacenters while fostering datacenters that are energy-efficient. Hence, providers running energy-efficient datacenters are able to offer cheaper prices to consumers, which consequently leads to a shift of workloads from energy-inefficient datacenters to energy-efficient datacenters.  The GreenCloud tax approach was implemented using the simulation environment CloudSim. We applied real data sets published in the SPEC benchmark for the executed simulation scenarios, which we used for evaluating the GreenCloud tax.
\end{abstract}

\section{Introduction and Motivation}

The scientific community proposed, over time, different visions of future cloud markets, 
identifying economic aspects even in Grids~\cite{weishaupl2005business,schikuta2005business,schikuta2008workflow}, the computing paradigm preceding Clouds. 
Trust and security are critical enablers for the digital economy because they foster confidence among participants, 
ensuring that individuals and businesses are willing to engage in digital transactions and share sensitive data. 
Without trust~\cite{weishaupl2006gset}, users would hesitate to adopt digital platforms, and without 
security~\cite{shaaban2019ontology}, the risk of fraud, data breaches, and cyberattacks would undermine the 
integrity of the ecosystem. 
Our past and ongoing work in the areas of services~\cite{vinek2011classification,mach2012generic} and 
workflow composition~\cite{kofler2010user,stuermer2009building} with a strong economic focus motivated our actual research on combining economic aspects and ecological cloud management, leading to the concept of a GreenCloud tax.

In recent years, the cloud market has increased rapidly, and today datacenters account for 2\% of the total energy consumption in the US~\cite{sverdlik_heres_2016}. Recently, Gartner~\cite{gartner017} forecasted a growth of 37\% for the infrastructure as a service (IaaS) market, 20\% for the software as a service (SaaS) market, and 23\% for the platform as a service (PaaS) market in 2017. Similar growth rates are expected for the next three years. This prediction is in line with Cisco's cloud forecast~\cite{cisco16}, which sees a significant growth of datacenters until 2020. The United States Data Center Energy Usage Report~\cite{shehabi_united_2016} introduces seven scenarios for predicting the energy consumption of datacenters in 2020. With the energy efficiency of 2010, the energy consumed by datacenters will more than triple from 70 billion kWh in 2015 to approximately 200 billion kWh in 2020. Currently, the report expects a moderate increment in energy consumption until 2020 by assuming significant improvements in energy efficiency. A historical trend from the Standard Performance Evaluation Corporation (SPEC)~\cite{spec16} shows that the recent efforts of industry and the scientific community in improving the energy efficiency of servers - a significant component of datacenters - were successful: Between the fourth quarter 2007 and the fourth quarter 2016, the server efficiency increased by a factor of approximately 17. 

Under the umbrella term Green Cloud computing, the scientific community introduces approaches and algorithms for boosting energy efficiency. Mastelic et. al.~\cite{mastelic_cloud_2015}  classified this field of research along four categories: network, server, cloud management system, and appliance. The scientific community develops novel or improved system and software architectures as well as algorithms - the focus is on enhancing energy efficiency by improving the underlying technology of datacenters. However,  Shehabi, one of the authors of the United States Data Center Energy Usage Report~\cite{shehabi_united_2016}, explained in~\cite{datacenterK16} that one main problem is that there are~\emph{millions} of inefficient datacenters, which are already running - the main challenge is to find ways to make them more efficient or to replace them, which accentuates the need for novel approaches. One first attempt for tackling this problem was recently introduced in~\cite{hinz_cost_2016} where the authors consider the CPU as the dominant consumer of energy in a datacenter. In order to reduce CPU and consequently energy consumption, it is recommended that cloud providers should charge consumers based on the consumed processing power, the CPU consumption. However, this pricing strategy applies to all datacenters and does not focus on the inefficient datacenters that are already running. Similarly, in~\cite{mach_consumer-provider_2011} and~\cite{NasirianiKW18}, the authors identified energy costs as the main cost factor of datacenters, which has to be considered for pricing the sold services.  In the paper at hand, we introduce an economical approach to tackle the described challenge. We developed and implemented a dynamic tax model for virtual machines (VMs) that we call GreenCloud tax. Thereby, virtual machines that are hosted by energy-efficient datacenters are lower taxed than virtual machines that are hosted by energy-inefficient datacenters. This gives providers that run energy-efficient datacenters a competitive advantage over providers that run inefficient datacenters. To the best of the authors' knowledge, no similar approach exists. In our previous work published in~\cite{pittl_cloudtax:_2016}, we introduced the idea of developing appropriate tax systems for cloud markets. The GreenCloud tax or similar ecological taxes, as well as their impacts, were not analyzed. 

\begin{figure}
 \begin{center}
    \includegraphics[width=0.55\linewidth]{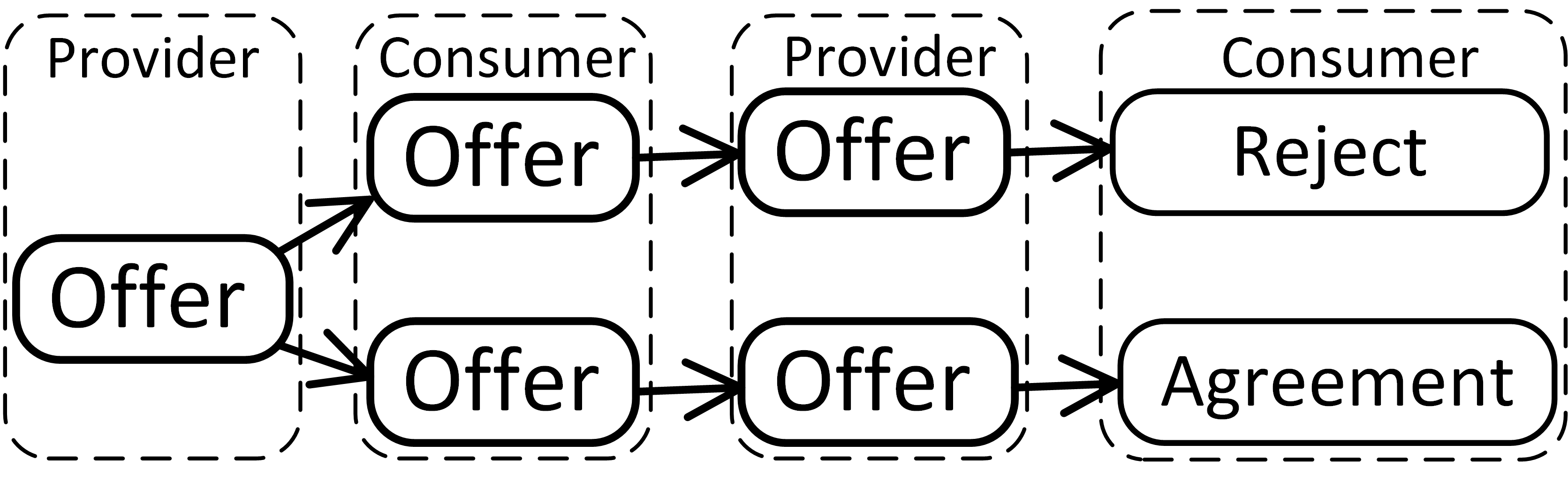}
    \caption{Exchange of offers and counteroffers during Bazaar negotiations}
    \label{fig:negotiationTree}
 \end{center}
\end{figure}

For the evaluation of the GreenCloud tax, we have to define the structure of the underlying cloud market. Currently, there is a shift from the classical supermarket approach - where consumers buy predefined virtual machines at fixed prices, such as on the Amazon EC2 on-demand market~\cite{amazon_amazon_nodate} - to more dynamic cloud markets~\cite{IrwinSSS17}. Amazon's EC2 spot market, where consumers bid for virtual machines, can be considered as the first step toward such a dynamic cloud market~\cite{amazon_amazon_nodate}. The scientific community introduced different visions of future cloud markets that are, e.g., based on centralized auctions~\cite{samimi_combinatorial_2016}, decentralized auctions~\cite{bonacquisto_strategy_2014}, or bilateral multi-round negotiations~\cite{pittl_classification16,dastjerdi_autonomous_2015}. For the evaluation of the GreenCloud tax, we assume that the market participants trade services using multi-round bilateral negotiations, aka Bazaar negotiations. These negotiations result in negotiation trees such as depicted in figure~\ref{fig:negotiationTree} and are in line with the vision of autonomous clouds, such as introduced in~\cite{HafezE16}. During such negotiations, the market participants use negotiation strategies that describe
\begin{inparaenum}[(i)]
\item how to rank received offers
\item how to respond to them and, if an offer is neither accepted nor rejected 
\item how to create counteroffers.  
\end{inparaenum}
As described in~\cite{pittl_classification16}, the scientific community developed different negotiation strategies, such as time-dependent strategies, where the duration of the negotiation determines, e.g., which counteroffers are created. The WS-Agreement Negotiation specification~\cite{waeldrich_ws-agreement_2011}, which was published by the Open Grid Forum, does not describe a concrete negotiation strategy. Instead, it fosters the development of Bazaar-based cloud markets by defining inter alia states of offers.


~

We extended CloudSim's~\emph {Bazaar-Component}~\cite{pittl_bazaar-extension:_2016} to simulate tax systems on Bazaar-based cloud markets. In the paper at hand, we used it for the evaluation of the GreenCloud tax; however, it is not limited to it. For running simulations with the extended Bazaar-Component, users first have to define the market participants and assign negotiation strategies to them. In the second step, the tax system has to be defined. Afterward, the simulation can be executed, and the results can be analyzed.


The main contributions of the paper are the following:
\begin{itemize}
	\item Design and development of a tax model fostering energy efficiency
	\item Implementation of the tax model on a CloudSim-based simulation environment
    \item A discussion and analysis of the results using real data sets from the SPEC benchmark~\cite{spec_server_nodate}.
\end{itemize}

The rest of the paper is structured as follows: In the following  section~\ref{sec:relatedWork}, existing approaches that foster the energy efficiency of the cloud are analyzed. The GreenCloud tax is introduced in section~\ref{sec:greencloudtax} while
the simulation environment - which is used for evaluating the GreenCloud tax  - as well as our simulation results, are introduced in section~\ref{sec:evaluation}. The paper closes with a conclusion and an outlook on our further research in section~\ref{sec:conclusion}.

\section{Related Work}
\label{sec:relatedWork}

This section is structured into two parts. First, approaches are introduced that optimize the technology of datacenters in order to improve energy efficiency. Second, we summarize approaches that apply economic approaches in order to enhance the energy efficiency of datacenters.


Approaches that improve the technology of datacenters in order to enhance their energy efficiency can be categorized into four groups~\cite{mastelic_cloud_2015}:
\begin{inparaenum}[(i)]
	\item Network: The traffic in networks is growing exponentially, so that they are a significant consumer of energy~\cite{kilper_power_2011}. All scientific efforts that reduce the consumption of energy in datacenter networks, networks between datacenters, and networks connecting end users with datacenters belong to this category. The scientific community, e.g., redesigns the hardware of network devices or improves the architecture of the networks. For example, in~\cite{KaidanAKKS17,KaidanKAK17} the authors identified a suboptimal distribution of the traffic load over optical network channels as a significant potential for improving energy efficiency. Different setups of mechanisms were developed in order to leverage this potential. 
	\item Server: This category includes all scientific efforts that optimize the energy consumption of racks, enclosures (cooling systems), and other devices, which do not belong to the previous category (networks). For example, novel cache strategies as well as the optimization of the heat load of, e.g., the CPU, belong to this category. For example, a cloud-based cooling solution was introduced in~\cite{VishwanathHB18}.
	\item Cloud Management System: Mastelic et.al~\cite{mastelic_cloud_2015} consider this category as a promising field of research. Virtual machine reconfiguration, virtual machine placement, virtual machine migration and consolidation, as well as virtual machine scheduling algorithms, virtualization software, and monitoring systems belong to this category. E.g. in~\cite{liu_greencloud:_2009} the authors reduce energy consumption by migrating virtual machines to other hosts while other researches such as~\cite{borgetto_energy-aware_2012} reduce energy consumption by shutting down idle hosts or by designing a lightweight cloud management system~\cite{ma_toward_2012}.  In~\cite{LiZZDT17}, a consolidation strategy for virtual resources was developed, which considers the tradeoff between energy consumption reduction and service level agreement compliance. A self-optimizing framework for cloud datacenters was introduced in~\cite{SinghCSB16}. It assigns the workloads to datacenter nodes considering energy efficiency as well as quality of service parameters. Similarly, in~\cite{OukfifBBB15} and~\cite{JinXLL17}, an energy-aware workflow optimization/scheduling algorithm was presented. An excellent paper was presented in~\cite{NasirianiKW18}, which aims at introducing tailored prices to consumers by considering~\emph{virtual battery management} on providers. Zhou et. al. use a peer-to-peer energy management approach for optimizing cloud energy storage~\cite{ZhouCLLY18}. Gai et. al. introduced a resource optimizing approach using heterogeneous cloud computing for cyber-physical systems~\cite{GaiQZS18}. The authors of~\cite{KhosraviAB17} presented an algorithm for VM placements in geographically distributed cloud datacenters.
	\item Appliance: In a perfect cloud system, only running applications consume energy~\cite{mastelic_cloud_2015}. However, usually, the runtime environments and operating systems are also significant consumers of energy. 
	Efficient processing paradigms such as MapReduce~\cite{dean_mapreduce:_2008} are an example of reducing energy consumption on the application level.  The malware detection and prevention system introduced in~\cite{MirzaMA16} also belongs to this category, which was developed with the aim of reducing energy consumption.
\end{inparaenum}


Compared to the approaches that try to improve energy efficiency by enhancing the underlying technology, approaches that try to apply economic principles to improve energy efficiency are rare. We see, e.g., the work published in~\cite{mach_consumer-provider_2011,mach_toward_2013} as part of this category, where the authors developed a cost model for cloud datacenters. Energy costs, together with the depreciation of datacenter infrastructure, are significant types of costs and hence they have a significant impact on the pricing decision of cloud providers running datacenters. The authors of~\cite{mohsenian-rad_energy-information_2010} show in their work that a huge number of parameters have to be considered for finding an optimal location of a datacenter. This problem was formalized as a non-linear programming problem, whereby the aim of the objective function is to minimize the carbon footprint and the energy consumption. Carbon taxes represent one parameter of this program, even if carbon taxes have only to be transferred by plants in the US today~\cite{mohsenian-rad_energy-information_2010}. Details of the referred carbon tax are not explained. Also, the authors of~\cite{yamini_cloud_2010} introduced an approach for finding the optimal location of datacenters in order to serve a maximum number of consumers. The author of~\cite{huu_auction-based_2013} suggests selling virtual machines for predefined time slots using auctions. Time slots for virtual machines in non-business hours will be cheaper than those that can be used during business hours. This gives consumers an incentive to use virtual machines during non-business hours, and therefore, the utilization of the datacenters gets more constant, leading to an optimized utilization of datacenter capacities. The authors of~\cite{deng2012adaptive} show in their paper that green datacenters should be able to mix~\emph{clean} and~\emph{dirty} energy sources in order to better fit consumer requirements and consequently to increase profit. Economical principles were also applied in the demand response domain. For example, in~\cite{liu2013data}, the authors introduced a demand response approach to save energy via workload shifting and local generation. Similarly, in~\cite{zhou2015smart}, a demand response approach was introduced by leveraging an auction mechanism. 

The related work analysis shows that the scientific community has developed different approaches and techniques to boost the energy efficiency of datacenters. However, we have not identified a tax-based approach that is used as leverage towards Green Cloud computing.

\section{Model of the GreenCloud Tax}
\label{sec:greencloudtax}

In literature such as~\cite{CelebicB15,mach_consumer-provider_2011,mach_toward_2013}, it is described that the costs of energy are a significant cost driver of cloud datacenters. However, inefficient datacenters with a high energy consumption are still running~\cite{datacenterK16}. This is because providers face inter alia a tradeoff between running costs such as energy costs and depreciation costs representing costs for acquisitions of new datacenter equipment. From an economic point of view, running old, inefficient datacenters can be a profit-maximizing strategy. For example, in cases in which the costs per consumed kWh are low, the investments in new, (the most) energy-efficient datacenter equipment will not be amortized within a short period, and so a rational provider will reject the investment - see~\cite{mach_toward_2013} for more information. The economic incentive to invest in energy-efficient equipment, such as energy-efficient servers, is too low for the provider running the inefficient datacenters. However, the usage of energy-efficient servers is an essential step towards the Green Cloud~\cite{elnozahy_energy-efficient_2002}. The GreenCloud tax - described in the following - stimulates investments in datacenters that improve energy efficiency.

Today, the value-added tax is used for taxing virtual machines. This tax is calculated based on the price. For example, the instance m4.3xlarge from Amazon's on-demand market (Windows with SQL, 8 vCPU, 26 ECU, 32GiB RAM, USA east (Ohio)) costs 1.728\$ per hour\footnote{price accessed on 28.10.2017} and consequently  41.472\$ for 24 hours. From the 41.472\$, Amazon receives 33.172\$, the tax authority receives the remaining 8.3\$ - assuming a tax rate of 20\%. 
Mankiw~\cite{mankiw_principles_2004} and other economic literature distinguish between the taxable base and the tax rate.  The value-added tax uses the price as a taxable base while the tax rate is proportional - in the previous case, the tax rate is 20\%. In the paper at hand, these two dimensions are analyzed for IaaS markets where virtual machines are traded and consequently taxed. Cloud services such as virtual machines are metered services - a key characteristic of cloud computing~\cite{mell_nist_2011} - so that the usage of alternative taxable bases - instead of the price - is feasible. Virtual machines can be taxed using one of their characteristics, such as processing power, RAM, or storage, which are the most important characteristics of a VM~\cite{pittl_cloudtax:_2016}. Using one of the aforementioned characteristics of a virtual machine as a taxable base makes this~\emph{resource} more expensive, and therefore consumers have an incentive not to use this resource. For example, processing power can be seen as a significant consumer of energy~\cite{hinz_cost_2016} that can be used as a taxable base. Thereby, each MIPS\footnote{ Millions Instructions Per Second (MIPS), abstract measure of processing power - see~\cite{buyya_cloudbus_2009} for more information} of a virtual machine is taxed with a certain amount of money. Consequently, virtual machines with a lot of processing power become more expensive while virtual machines with less processing power get cheaper. Thus, some consumers may switch to VMs with less processing power, leading to a total reduction of the consumed processing power. Similarly, each GB of RAM cloud be taxed as well as each GB of storage. Table~\ref{tab:taxes} summarizes the taxable bases of virtual machines. Taxes using the price as a taxable base are called~\emph{Price-based tax} while taxes using one of the other characteristics of VMs as a taxable base are called~\emph{Resource-based taxes}. The~\emph{Fee} and the~\emph{GreenCloud tax} are exceptions, which are discussed in the following.

The tax rate defines how many percent of the total price of the virtual machine are taxed, as the following equation shows. 
\begin{equation}	
	\label{equ:taxrate}
	\text{tax rate}=\frac{\text{tax}}{\text{price of virtual machine}}
\end{equation}
Mankiw~\cite{mankiw_principles_2004} distinguishes between a proportional tax rate, a progressive tax rate, a regressive tax rate, and a lump sum tax rate. The previously described value-added tax is an example of a proportional tax rate. It is, e.g., 20\% for cheap virtual machines as well as for expensive virtual machines. On the contrary, the progressive tax rate increases with an increasing taxable base. Usually, a progressive tax rate is used for payroll. Persons with a high payroll face a higher tax rate than persons with a low payroll. A progressive tax rate for virtual machines would imply that expensive virtual machines are taxed with a higher tax rate than cheap virtual machines, which are taxed with a lower tax rate - using the price as a taxable base. Consider the following example of a progressive tax rate: a virtual machine with a price of 200\$ is taxed with a tax rate of 20\% while a virtual machine with a price of 100\$ is taxed with a tax rate of 5\%. Regressive tax rates are inverse to progressive tax rates. Here, the rate decreases with an increasing taxable base. The lump sum tax is completely independent of the taxable base and is a special form of the regressive tax rate. Hence, it is comparable to a fee - we used this term for the described tax models in table~\ref{tab:taxes}.

\begin{table*}
\caption{Taxable bases and tax rates of virtual machines}
\begin{center}
\scriptsize
\begin{tabular}{lcccc}
\hline
\hline
Taxable base/Tax rate & \textbf{Lump Sum} & \textbf{Progressive} & \textbf{Regressive} & \textbf{Proportional} \\
\hline
\hline
Price & Fee & Price-based Tax & Price-based Tax & Price-based Tax \\
Storage & Fee & Resource-based Tax  & Resource-based Tax & Resource-based Tax \\
RAM & Fee & Resource-based Tax  & Resource-based Tax & Resource-based Tax  \\
Processing Power  & Fee  & Resource-based Tax  & Resource-based Tax  & Resource-based Tax \\

\hline
Server energy efficiency & Fee & GreenCloud Tax & - & GreenCloud Tax \\
\hline
\end{tabular}
\end{center}
\label{tab:taxes}
\end{table*}

The cost models presented in~\cite{mach_consumer-provider_2011} and~\cite{mach_toward_2013} identified all resources, such as processing power, RAM, and storage of virtual machines, as significant consumers of energy. Using one or several of these resources as a taxable base makes them more expensive. Thus, consumers have an incentive not to use the resources that are used as the taxable base. So the consumption of the resources, including the corresponding energy consumption, will be reduced. Such a resource-based tax does not consider the energy efficiency of the servers\footnote{in this paper, we use the term server broadly: it represents processing power, RAM, and storage} that host the virtual machines. According to such resource-based tax models, a virtual machine running on an energy-efficient server will be taxed with the same amount as the identical virtual machine that runs on an energy-inefficient server. So the resource-based taxes do not set stimuli for consumers to buy virtual machines from providers that run energy-efficient servers. Moreover, the provider using the energy-efficient server has no competitive advantage - from the tax authorities' point of view - over the provider that runs energy-inefficient servers. Therefore, we introduce the GreenCloud tax as shown in the last row of table~\ref{tab:taxes}. It is a tax model that uses the energy efficiency of the server, which hosts the virtual machine, as a taxable base. The effects of such a tax model are twofold: It gives consumers an incentive to switch to providers that host VMs on energy-efficient servers, as their taxes are lower. Similarly, providers have an incentive to invest in energy-efficient servers in order to offer more attractive prices to consumers because of the reduced tax rate - using the GreenCloud tax - and the lower energy consumption costs. As table~\ref{tab:taxes} shows, the GreenCloud tax can use different tax rates. A proportional tax rate implies that the virtual machine is taxed proportionally to the energy efficiency of the servers hosting the VM - we used such a tax rate in our use case presented in section~\ref{sec:evaluation}. A progressive tax rate implies a disproportionately higher tax rate for virtual machines that are hosted by energy-inefficient servers. In table~\ref{tab:taxes}, the combination with the regressive tax rate is missing. It would imply that virtual machines that run on energy-inefficient servers face a lower tax rate than those running on energy-efficient servers, which is contradictory to the goal of the GreenCloud tax.

The GreenCloud tax - as described before - neither considers the price nor the resources of a virtual machine for the calculation of the tax. Hence, large and expensive virtual machines, which are e.g. used as database servers, are taxed with the same amount as small and cheap virtual machines, which are e.g. used as a workstation if they are hosted by the same server. Alternatively, combined tax bases can be used instead of a pure GreenCloud tax. The following equation represents a GreenCloud tax using a combined taxable base. Here, the price as well as the energy efficiency of the server hosting the VM are used as a taxable base. 
\begin{equation}
	\label{equ:energytax}
	\text{tax}=\text{price} \cdot \text{tax rate} \cdot \text{efficiency factor}
\end{equation}
The first two terms ($\text{price} \cdot \text{tax rate}$) calculate a price-based tax that is multiplied by an $\text{efficiency factor}$. This factor represents the energy efficiency of the server that hosts the virtual machine. It is low if the server is energy-efficient, but it is high if the server is energy-inefficient. Determining the energy efficiency factor is one of the main challenges of the GreenCloud tax. Firstly, profiling servers along energy efficiency is not trivial. Secondly,   migrations of VMs between different servers as well as idle times make the calculation of the $\text{efficiency factor}$ difficult. We see this issue as part of our further research. In the paper at hand, we used data from the SPEC benchmark published in~\cite{spec_server_nodate} for the calculation of the $\text{efficiency factor}$. Thereby, so-called ssj\_ops/watt\footnote{The more ssj\_ops the system under test can produce with one watt of power, the better is the efficiency of the system under study - see https://www.spec.org/power/docs/SPECpower\_ssj2008-User\_Guide.pdf} are used as a key metric for comparing the energy efficiency of servers. The metric ssj\_ops/watt is used by the SPEC benchmark to compare the energy efficiency of servers, while the aforementioned MIPS are used by CloudSim as a metric to compare the processing power capacity of servers.

\begin{figure}
		\begin{subfigure}[b]{0.49\linewidth}
			\centering
			\includegraphics[width=0.97\linewidth]{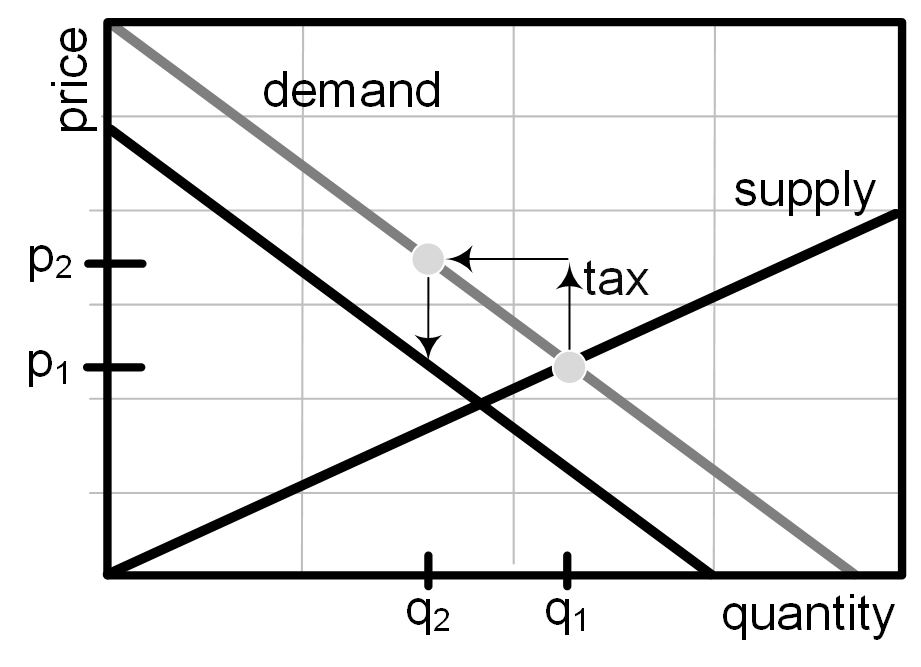}
			\caption{Shift of demand curve}
			\label{fig:taxSystem}
		\end{subfigure}
		\begin{subfigure}[b]{0.5\linewidth}
			\centering
			\includegraphics[width=1\linewidth]{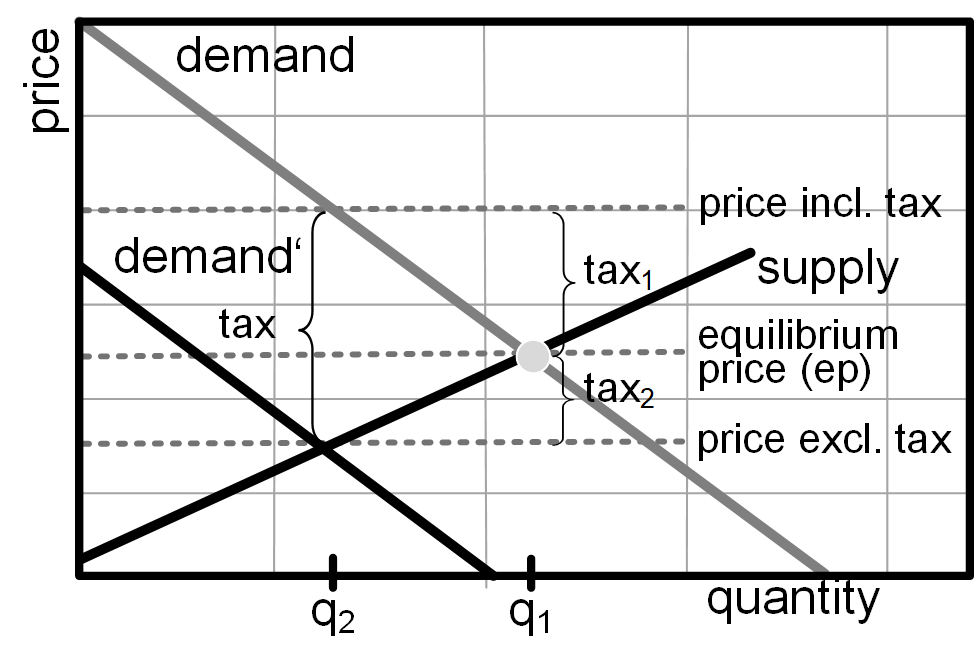}
			\caption{Tax incidence}
			\label{fig:taxImpact}
		\end{subfigure}
		\caption{Impact of taxes on demand and supply}
	\end{figure}

The impact of taxes is visualized in  figure~\ref{fig:taxSystem}. It shows a typical market with a demand and supply curve. Virtual machines can be considered as virtual goods that are supplied by providers and demanded by consumers. Therefore, fundamental market mechanisms can be applied to VMs. The initial demand curve is the gray one. The price $p_{1}$ of the good traded on the market is determined by the intersection of the demand curve and the supply curve. At this price, the consumers demand quantity $q_{1}$. Assume that the consumer has to transfer the tax, a fixed amount of money, in addition to the price $p_{1}$. Hence, the consumer demands as much as $p_{1}$ plus the tax. So the demand curve shifts inwards (black curve), which represents the demand curve including taxes.




According to the tax incidence theory, the entity that transfers taxes does not necessarily pay the taxes - it does not receive the total tax burden~\cite{mankiw_principles_2004}. This is illustrated in figure~\ref{fig:taxImpact}. Again, this figure shows two demand curves: the gray demand curve shows the demand curve before the introduction of the tax, while the black demand curve is the demand curve after the introduction of the tax. The intersection of the demand and the supply curve forms the so-called equilibrium price. After introducing the tax, the demand curve shifts inwards, leading to a new equilibrium price ($\text{price excl. tax}$) as well as to a new quantity ($q_{2}$). Even if the consumer has to transfer the tax, both the consumer and the provider have to pay a tax as described in the following:


\begin{itemize}
	\item Before the tax is introduced, the consumer pays the equilibrium price to the provider. After the consumer pays the lower price excluding tax, there is additionally the tax. So finally, the consumer pays the price, including the tax, for the good traded on the market. The difference between the initial equilibrium price and the price including tax is represented by $tax_{1}$.
	\item The provider gets the equilibrium price before the tax is introduced. After, the equilibrium price drops and the provider receives only the price, termed price excluding tax in figure~\ref{fig:taxImpact}. The difference between the initial price and the price excluding tax is represented by $tax_{2}$.
\end{itemize}

%
%

\begin{figure}
		\begin{subfigure}[b]{0.60\linewidth}
			\centering
			\includegraphics[width=0.87\textwidth]{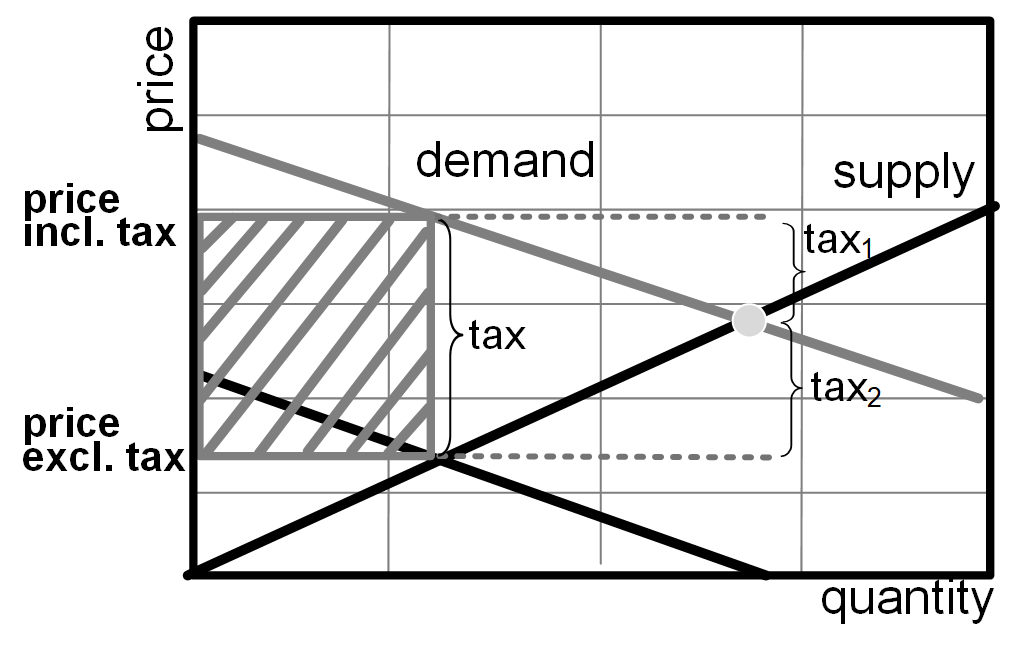}
			\caption{Elastic demand}
			\label{fig:elasticExample}
		\end{subfigure}
		\begin{subfigure}[b]{0.38\linewidth}
			\centering
			\includegraphics[width=0.94\textwidth]{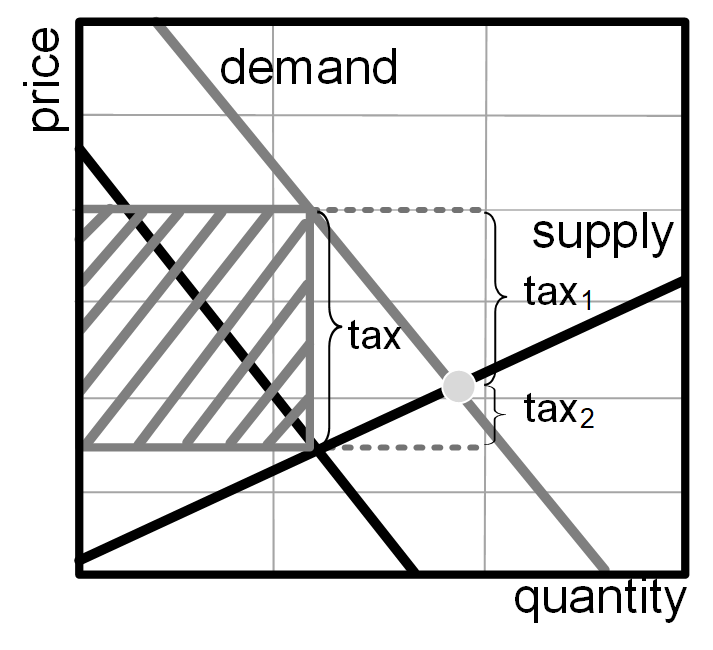}
			\caption{Inelastic demand}
			\label{fig:elasticExample2}
		\end{subfigure}
		\caption{Tax burden examples}
		\label{fig:elasticExamples}
	\end{figure}

$tax_{1}$ and $tax_{2}$ form together the total tax determined by the tax authority.
Again, the example illustrates that the entity that transfers the tax does not necessarily pay it. Instead, the price elasticity of the demand and supply curve determines the amount of the tax an entity has to pay as defined by the following equation:
\begin{equation}
	\text{price elasticity}=\Big|\frac{dQ/Q}{dP/P}\Big|
\end{equation}
The higher the price elasticity of the demand or supply curve, the lower the tax the consumer or provider has to pay. 
In both examples shown in figure~\ref{fig:elasticExamples}, the tax is transferred by the consumer, and its amount is identical. In figure~\ref{fig:elasticExample}, the demand is elastic so that the provider has to pay a larger share of the tax than the consumer. In figure~\ref{fig:elasticExample2}, the demand is inelastic so that the consumer pays most of the tax. 
%
In both examples, the quantity sold is lower than in the situation without taxes. The tax revenue is represented by the dashed areas in the figures.
It is calculated by multiplying the quantity by the tax amount, as shown in the equation:
\begin{equation}
	\text{tax revenue}=\text{quantity} \cdot \text{tax size}
\end{equation}
Increasing taxes makes goods more expensive, leading to a reduced quantity of traded goods. The resulting two effects are described in the following:
\begin{inparaenum}[(i)]
	\item \emph{Effect 1:}	
	By increasing the size of the tax, the tax revenue increases by each item sold.
	\item \emph{Effect 2:}
	Increasing the size of the tax leads to reduced quantity because goods get more expensive. Consumers who have a lower willingness to pay than the price do not purchase the good anymore. So some goods will not be sold, and consequently, no tax revenue is earned.
\end{inparaenum}

The Laffer Curve~\cite{mankiw_principles_2004} depicted in figure~\ref{fig:lafferCurve} visualizes these two effects. If the tax size is low, then increasing the tax size increases the tax revenue because Effect 1 dominates Effect 2. On the contrary, if the tax size is already very high, increasing the tax size decreases the tax revenue. 
Market participants leave the market due to the high tax, and so the number of transactions decreases. Therefore, Effect 2 dominates Effect 1.

\begin{figure*}[htbp]
 \begin{center}
    \includegraphics[width=0.95\linewidth]{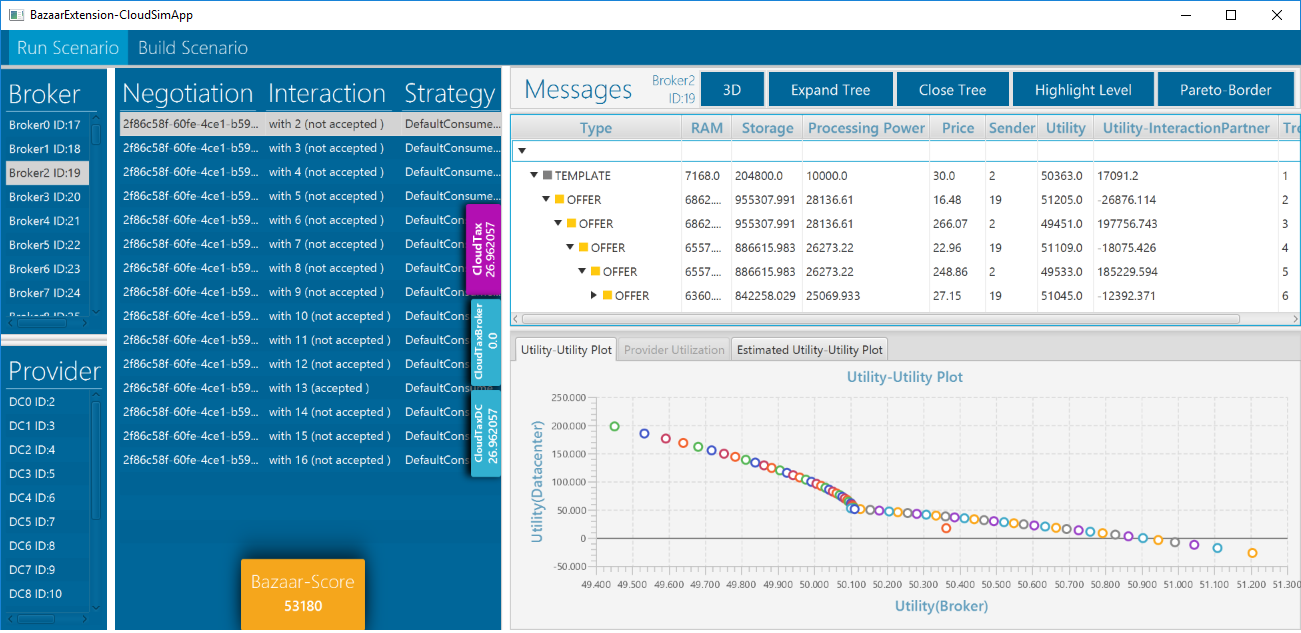}
    \caption{Simulation environment for simulating taxes such as the GreenCloud tax}
    \label{fig:screenshot}
 \end{center}
\end{figure*}

\section{Evaluation of the GreenCloud Tax}
\label{sec:evaluation}

We extended CloudSim's Bazaar-Extension~\cite{pittl_bazaar-extension:_2016} for evaluating the GreenCloud tax. Before running simulations with our simulation environment, you have to define the market participants, including their negotiation strategy and the GreenCloud tax that should be used. After running the simulation, you can analyze the resulting resource allocation. Figure~\ref{fig:screenshot} shows a screenshot of the simulation environment. On the left side, the market participants are listed who attend the simulation scenario (the broker represents consumers). After selecting one of the market participants, you see its corresponding negotiations in the second column. In the shown screenshot, broker $2$ negotiated with $15$ providers. By selecting a negotiation, its exchanged offers - which contain a description of a virtual machine - are visualized as a tree list, as the right side of the figure shows. Each offer of this tree list is visualized as a dot on the utility-utility plot. The utility values are calculated using utility functions - market participants use them for ranking offers - see~\cite{pittl_negotiation-based_2015} for more information. The ordinate of the plot shows the utility of the virtual machines contained in offers for the provider (datacenter), while the abscissa shows the utility of the virtual machine contained in the offer for the consumer. The tax revenue gained by the tax authority of the executed market scenario is shown in the violate box.

\begin{figure}
 \begin{center}
    \includegraphics[width=0.4\linewidth]{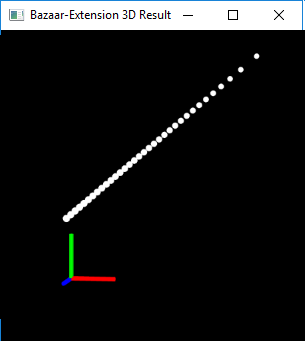}
    \caption{3D visualization of the offers exchanged during negotiation (white dots represent the offers, the axes represent the VM characteristics)}
    \label{fig:3dplot}
 \end{center}
\end{figure}

The simulation environment supports a 3D visualization of the offers that are exchanged during a selected negotiation, as figure~\ref{fig:3dplot} shows\footnote{the library F(X)yz3D was used for creating this component}. The blue axis represents the storage, the red axis represents the processing power, while the green axis represents the RAM. The white dots represent the virtual machines that are contained in the offers. The used negotiation strategies determine the creation of virtual machines, which are offered to the negotiation partner and consequently the position of the white dots in figure~\ref{fig:3dplot}.

\begin{figure}
     \begin{subfigure}{0.46\linewidth}
        \includegraphics[width=1\linewidth]{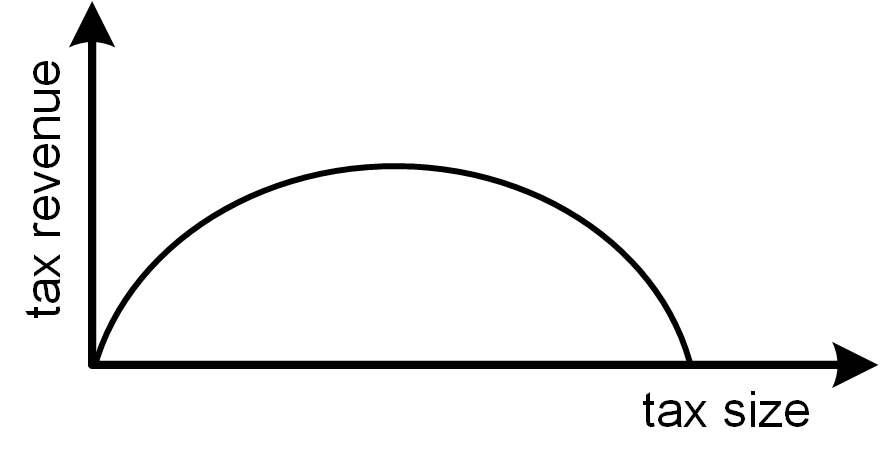}
        \caption{Laffer Curve}
        \label{fig:lafferCurve}
    \end{subfigure}
    \begin{subfigure}{0.53\linewidth}
        \includegraphics[width=1\linewidth]{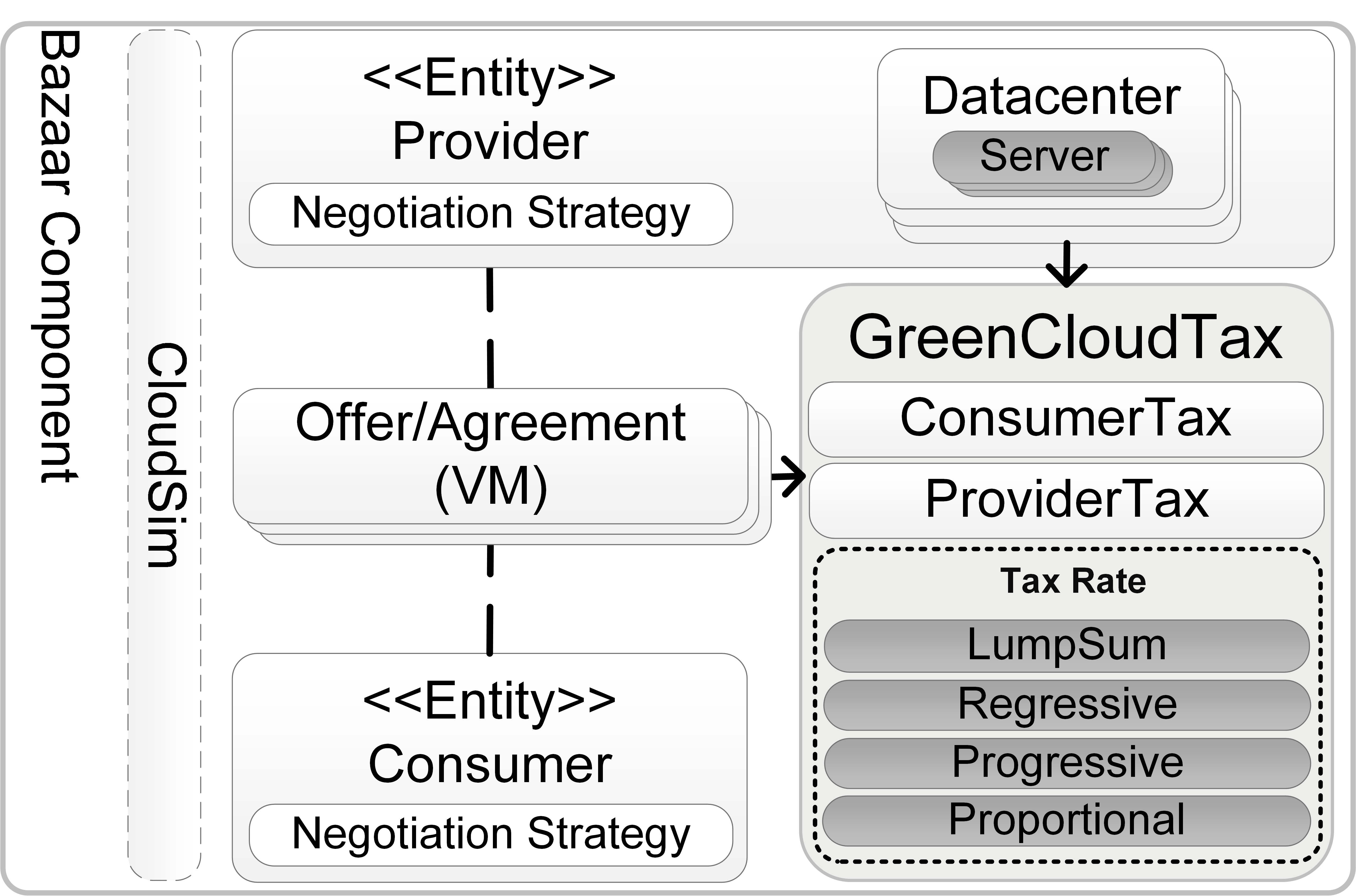}
        \caption{Overview of the architecture of the GreenCloudTax component}
        \label{fig:architecture}
    \end{subfigure}
    \caption{Laffer Curve and overview of the architecture of the GreenCloudTax component}
\end{figure}

A high-level architecture of the developed tax simulation component is illustrated in figure~\ref{fig:architecture}. The GreenCloudTax component, which we developed, is queried during negotiation by consumers as well as by providers and returns the estimated tax by evaluating the energy efficiency of the server where the VM is planned to run. If an agreement is formed between two market participants, then it is passed to the GreenCloudTax component for calculating the tax revenue. Different tax rates are supported. 

In the presented use case, CloudSim's datacenter simulation capabilities were used. CloudSim is able to simulate the utilization of datacenters as well as VM migrations and placement - see~\cite{buyya_modeling_2009} for more information. We extended it and provided an inter alia negotiation simulation before VMs can be placed on a provider's datacenter, as well as a simulation of tax models.

\subsection{GreenCloud Tax Use Case}

We created a market scenario with $60$ consumers and $15$ providers, which run a datacenter and are able to host $10$ consumers. So all providers have the same capacity but use different types of servers to ensure comparability. Further, we assumed typical configurations such as homogeneous datacenters. This means that a datacenter uses only one type of server. Table~\ref{tab:servers} shows the types of servers used in the datacenters. The datacenter of the first provider consists of HP ProLiant DL385 servers, while the datacenter of the second provider consists of Acer Altos R520 servers. Provider 15 runs its datacenter with UniServer R49500 servers. 
The market participants trade services using Bazaar negotiations. Therefore, we implemented the time-dependent negotiation strategy introduced in~\cite{dastjerdi_autonomous_2012,dastjerdi_autonomous_2015}. 


\subsection{Consumer Strategy}

According to~\cite{dastjerdi_autonomous_2015}, consumers have a maximum as well as a minimum value for each characteristic of the traded good. Thus, in the case of virtual machines, consumers have maximum and minimum values for processing power, storage, RAM, and price. Bilateral negotiation strategies have to describe
\begin{inparaenum}[(i)]
    \item which offers are accepted, and
    \item if the offers are not accepted, how are counteroffers created?
\end{inparaenum}
For better readability of this paper, we start with the description of the creation of counteroffers.

\emph{Creation of counteroffers}: Counteroffers are denoted with $O^{t}_{a\rightarrow b}$ whereby $a$ is the sender and $b$ is the receiver of the offer. $t$ represents the time at which the offers were sent from $a$ to $b$. In~\cite{dastjerdi_autonomous_2015}, the following strategy is suggested for creating counteroffers:
    \begin{equation}
    \label{equ:offergeneration}
        O^{t}_{a\rightarrow b}[i]=
           \begin{cases}
      min_{i}^{a}+\alpha_{i}^{a}(t) \cdot (max_{i}^{a}-min_{i}^{a}) &  \text{if } V_{i}^{a} \text{ decreasing} \\
      min_{i}^{a}+(1-\alpha_{i}^{a}(t)) \cdot  (max_{i}^{a}-min_{i}^{a}) &  \text{if } V_{i}^{a} \text{ increasing} \\
   \end{cases}
    \end{equation}
    $i$ is a characteristic of the virtual machine, $V_{i}^{a}$ is the value of characteristic $i$ for sender $a$, $max_{i}$ and $min_{i}$ are the minimum and the maximum values for characteristic $i$ and $\alpha_{i}^{a}(t)$ is a time-dependent variable that takes on values between $0$ and $1$. According to the strategy, the consumers start with offers that maximize their utility. In cases of virtual machines, the initial offer will contain the following values: $max_{RAM}^{a}$, $max_{storage}^{a}$, $max_{\text{processing power}}^{a}$, $min_{price}^{a}$. Over time, the characteristics are modified until the deadline is reached. So the last offer of the consumer (if an agreement is not formed before) is: $min_{RAM}^{a}$, $min_{storage}^{a}$, $min_{\text{processing power}}^{a}$ $\text{ and }$ $max_{price}^{a}$. An example where consumers use this strategy is shown in the utility-utility plot in figure~\ref{fig:screenshot}. $\alpha_{i}^{a}(t)$ determines how fast the initial $max$/$min$ values are decreased/increased to final $min$/$max$ values. Therefore~\cite{dastjerdi_autonomous_2015} suggested to use, e.g., exponential functions. We used the polynomial function as described in the following equation:
    \begin{equation}
        \alpha_{i}^{a}(t)=k_{i}^{a}+(1-k_{i}^{a}) \cdot \Big(\frac{min(t,t_{max})}{{t_{max}}^{1/\beta}}\Big)
    \end{equation}
    $t_{max}$ represents the deadline. For the following use case, $\beta=2$ and $k=0$ were used.

\emph{Offer acceptance conditions}:
    Inspired by the offer acceptance conditions described in~\cite{dastjerdi_autonomous_2015} consumers accept offers if the following condition is fulfilled:  $UV_{O^{t}_{b\rightarrow a}}>UV_{O^{t+\epsilon}_{a\rightarrow b}}$.
    So a received offer $UV_{O^{t}_{b\rightarrow a}}$ is accepted by consumer $a$ if the utility of the received offer exceeds the utility of the counteroffer, which would be created in response to the received offer according to equation~\ref{equ:offergeneration}. For applying the decision rule, a definition of a consumer utility function for calculating the utility is necessary. In~\cite{dastjerdi_autonomous_2012}, a consumer utility function is not introduced - hence we used the following one, inspired by~\cite{pittl_negotiation-based_2015}, which considers basic economic principles:
    \begin{equation}
     \begin{aligned}		
        UV_{con.}=  log(storage \cdot w_{storage})+     log(processing p. \cdot  
         w_{\text{processing p.}})+ \\ log(RAM   
         \cdot w_{RAM})+ 
         log(max_{price}-price) \cdot          w_{price}
     \end{aligned}
    \end{equation}
\begin{table*}
\caption{Simulation parameters}
\begin{center}
\begin{tabular}{lclc}
\hline
\hline
\textbf{Consumer} & & &  \\
Parameter & Value & Parameter & Value \\
\hline
$w_{RAM}$ & 0.01 & $w_{\text{processing power}}$ & 0.01 \\
$w_{storage}$ & 0.01 & $w_{price}$ & 0.97 \\
$min_{RAM}$ & 3072 MB & $max_{RAM}$ & 7168 MB \\
$min_{\text{processing power}}$ & 5000 MIPS & $max_{\text{processing power}}$ & 30000 MIPS\\
$min_{storage}$ & 102400 MB & $max_{storage}$ & 1024000 MB\\
$min_{price}$  & 10\$ & $max_{price}$  & 23\$-100\$  \\
$t_{max}$  (simulation clock)   &  7200  &  & \\
\hline
\hline
\textbf{Provider} & & &  \\
Parameter & Value & Parameter & Value \\
\hline

$A_{RAM}$ &  0.8 & $A_{\text{processing power}}$ & 0.8\\
$A_{storage}$ & 0.8 & $w_{RAM}$ & 0.5 \\
$w_{storage}$ & 0.25 & $w_{\text{processing power}}$ & 0.25\\
$\text{MinRP}_{storage}$ & 0.000002\$- 0.0000022\$& $\text{MaxRP}_{storage}$ & 0.00001\$- 0.000011\$ \\
$\text{MinRP}_{RAM}$     & 0.002\$-0.0022\$ & $\text{MaxRP}_{RAM}$ &  0.03\$-0,033\$ \\
$\text{MinRP}_{\text{processing power}}$ & 0.0002\$-0.00022\$ & $\text{MaxRP}_{\text{processing power}}$ & 0.001\$-0.0011\$ \\
$t_{max}$  (simulation clock)   &  7200  &  &  \\
\hline
\end{tabular}
\end{center}
\label{tab:parameters}
\end{table*}

\subsection{Provider Strategy}

The provider strategy for the creation of counteroffers is similar to the consumer's strategy.
In~\cite{dastjerdi_autonomous_2015}, the provider strategy is responsible for suggesting a price for received offers. Therefore, the provider calculates so-called resource prices $RP_{it}$ for each characteristic $i$ of the virtual machine (RAM, storage, processing power) at time $t$. The resource prices are time-dependent, as shown in the following. The provider has for each characteristic a maximum resource price ($MaxRP$) as well as a minimum resource price ($MinRP$). The structure of this equation is similar to the one in equation~\ref{equ:offergeneration}.
    \begin{equation}
        RP_{it}=Min RP_{i}+\alpha RP_{i}(t)(Max RP_{i}- Min RP_{i})
    \end{equation}
$\alpha RP_{i}(t)$ is a time-dependent factor taking values between $0$ and $1$. The authors of~\cite{dastjerdi_autonomous_2015} suggest using a polynomial function for calculating this factor as defined in the following:
    \begin{equation}
        \alpha RP_{it}=IRP_{i}+(1-IRP_{i}) \Big(\frac{min(t,t_{max})}{t_{max}}\Big)^{1/\beta_{i}}
    \end{equation}
 $IRP$ stands for initial resource price. We used the $MaxRP$ as  $IRP$. It can be distinguished between two $\beta$ values: The first one is called resource-aware $\beta$, while the second one is called priority-oriented $\beta$. For the resource aware $\beta$ the share of the available resource $A_{i}$ for the resource characteristic $i$ has to be calculated so that we can calculate the average share of available resources: $\bar{A}=\frac{\sum_{i=1}^{m} A_{i}}{m}$.
The resource aware $\beta$ is calculated as $\beta_{i}=e^{A_{i}-\bar{A}}$.
In the executed use cases, we assumed an equal utilization of all resources. The preference-based $\beta$ is calculated as follows: $\beta_{i}=e^{1/n-w_{i}}$, whereby $n$ represents the number of resources and $w_{i}$ represents the importance factor so that $\sum_{i=1}^m w_{i}=1$.

The resource prices are calculated twice: once using the resource-aware $\beta$ and once using the preference-based $\beta$. The two prices are combined for each resource using a weighted average:
 \begin{equation}
    RP_{it}=RP_{it}^{\text{resource aware } \beta} \cdot 0.5 + RP_{it}^{\text{preference } \beta} \cdot 0.5
\end{equation}
 Finally, the resource prices are summarized to a final price $P_{t}=\sum_{i=1}^m RP_{it} \cdot i$.
 Consumers either accept the price suggested by the provider and form a binding agreement or respond with counteroffers. A decision rule, which describes when to accept an offer from a consumer, is missing, and so only the consumer accepts offers. Table~\ref{tab:parameters} summarizes the remaining parameters.  The parameters were chosen so that a consumer is charged a price comparable to Amazon's prices. E.g., the $30^{th}$ consumer wants at time $t$ a virtual machine with 4 GB RAM, 10000 MIPS processing power, and 300 GB storage.  This virtual machine is comparable to e.g., Amazon's instance type~\emph{t2.medium} (4 GB RAM, low processing power).   On the on-demand market, you pay for a t2.medium instance approximately 46\$   for a month (availability zone Ohio, Windows operating system).  Here, we have to stress that different parameters exist that have a significant effect on the price. The prices of the providers were chosen so that the average providers (see the following section) charge at $t_{max}/2$ comparable prices for such a virtual machine. Amazon sells storage separately via its EBS platform, where a monthly fee per used GB has to be paid - the provider's prices for storage were defined similarly to the prices for processing power and RAM, so that they are comparable to Amazon's prices at time $t_{max}/2$.


\subsection{Simulation Setup}

We simulated an idealized market as illustrated in figure~\ref{fig:elasticExamples}. The demand curve represents the willingness to pay of consumers who are participating in the market. Consumers on the left side of the curve have a high willingness to pay, while consumers on the right side of the curve have a low willingness to pay. This is reflected by the maximum prices which are given in table~\ref{tab:parameters}. Similarly, the supply curve represents the costs of the providers. The right side of the supply curve represents providers that have high costs, while the left side of the supply curve represents providers that have low costs. This is reflected by the resource prices, which are described in table~\ref{tab:parameters}. Energy-efficient servers usually use the latest technology and, therefore, are more expensive than servers with lower energy efficiency. Hence, we assume that providers running their datacenter with energy-efficient servers have higher costs than providers running their datacenters with energy-inefficient servers - the reduced energy costs of efficient servers can not compensate for the higher acquisition costs in the observed period of time. This is in line with the reality where inefficient servers are still running~\cite{datacenterK16}. We assigned the resource prices to providers proportional to their efficiency. So the first provider $P1$, which runs the most inefficient servers (HP ProLiant DL385 in our case), has the lowest resource prices, while the provider $P15$, which runs the most efficient servers (H3C UniServer R4900), has the highest resource prices.


With this setup, we executed one scenario using the widely used value-added tax model and eleven scenarios using the GreenCloud tax. Within the description of equation~\ref{equ:energytax}, we stated that the efficiency factor is high if the server hosting the virtual machines is not energy-efficient. In our use cases, we calculated the $\text{efficiency factor}$ of a server $i$ as follows. 
\begin{equation}
			\text{efficiency factor}_{i}=  (1-\text{interpolation factor}_{i})
				  \cdot \text{eco penalty}
\end{equation}
Thereby, the interpolation factor is $1$ for the most energy-efficient servers - the last server shown in table~\ref{tab:servers} - and $0$ for the most energy-inefficient server - the first server shown in table~\ref{tab:servers}. The equation for calculating the interpolation factor of a server $i$ is shown in the following. We calculated it by normalizing the ssj\_ops/watt of server $i$. 


	

\begin{equation}
 \text{interpolation} \text{ factor}_{i}  =  
 \frac{ssj\_ops/watt_{i}- ssj\_ops/watt_{\text{minimum}}}{ssj\_ops/watt_{\text{maximum}}-ssj\_ops/watt_{\text{minimum}}} 
\end{equation}
$ssj\_ops/watt_{minimum}$ represents the $ssj\_ops/watt$ of the most inefficient server while $ssj\_ops/watt_{maximum}$ represents the $ssj\_ops/watt$ of the most efficient server. The $\text{eco penalty}$ is defined by the tax authority. The higher it is, the higher the tax that the servers - except the most efficient server - have to pay. In the eleven scenarios using the GreenCloud tax, we varied the $\text{eco penalties}$.

		 

\subsection{Simulation Results}



\begin{figure*}
 \begin{center}
    \includegraphics[width=1\linewidth]{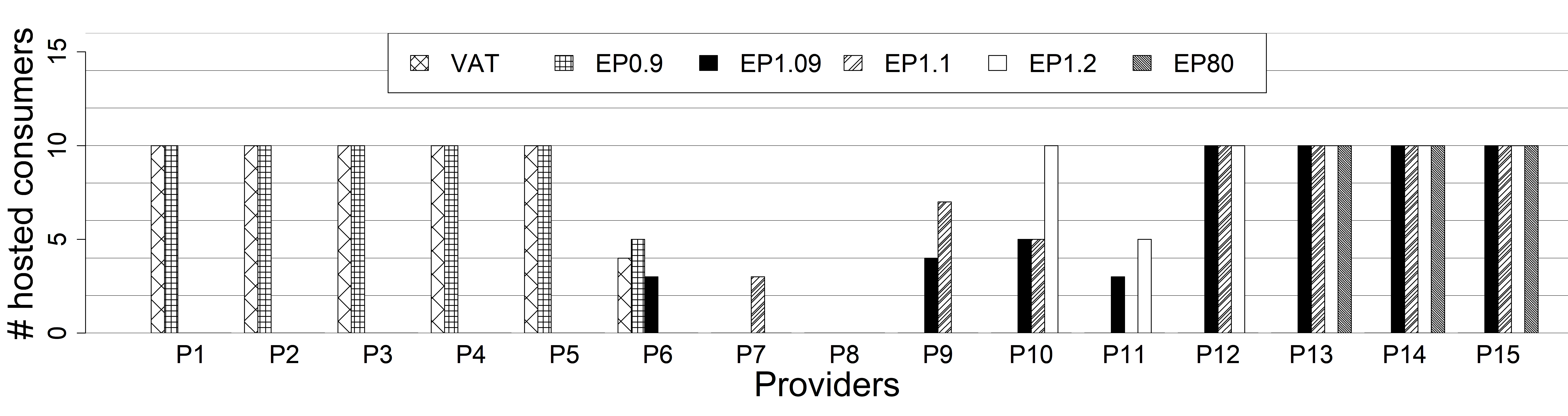}
    \caption{Simulation results - number of consumers hosted by the providers using different tax systems (10 is the capacity limit), VAT=value-added tax, EP=eco penalty for the used GreenCloud tax}
    \label{fig:scenarioResults}
 \end{center}
\end{figure*}

\begin{figure}
 \begin{center}
    \includegraphics[width=0.6\linewidth]{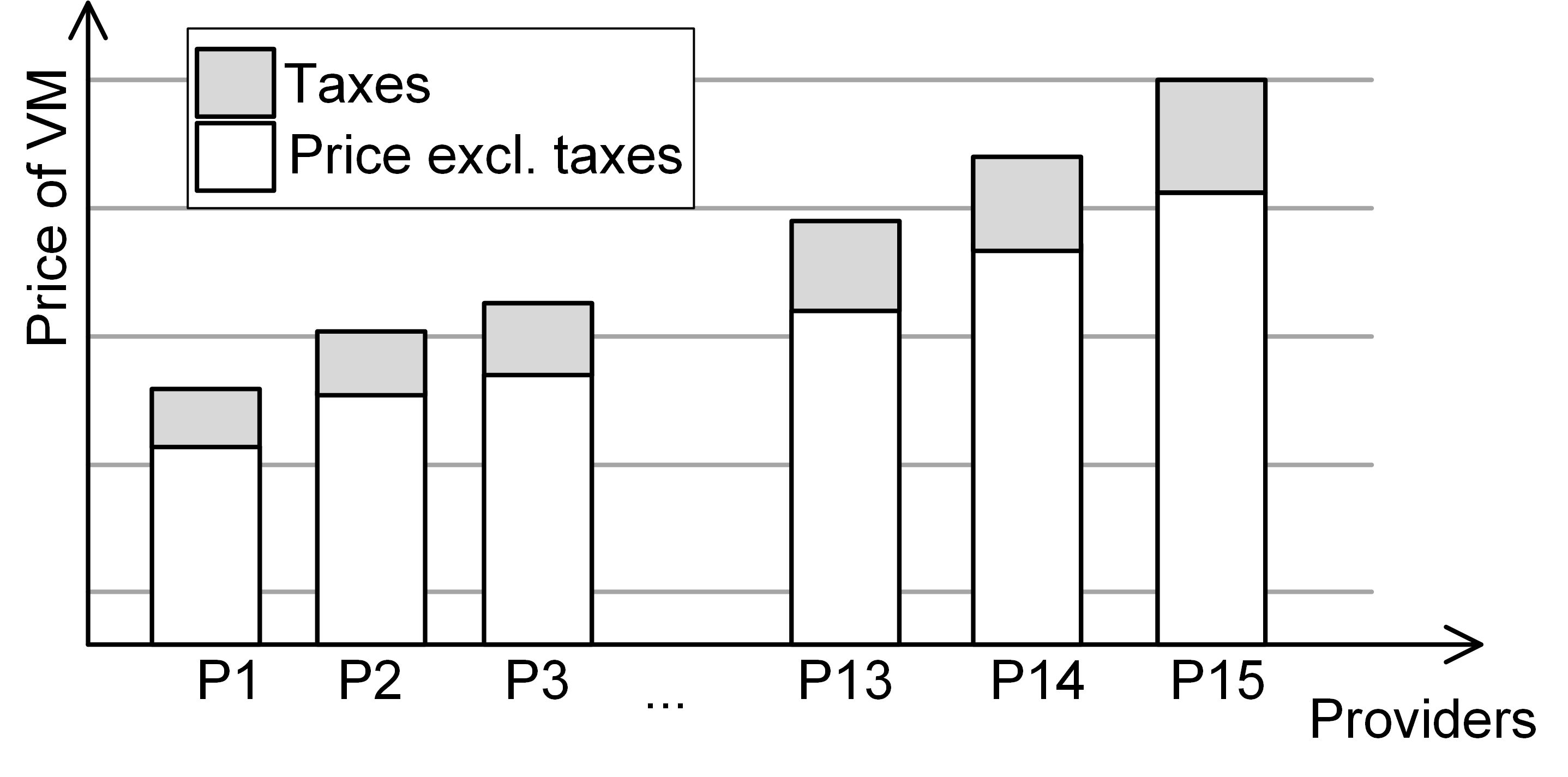}
    \caption{Schematic overview of the prices of a virtual machine using the value-added tax - tax is proportional to the price}
    \label{fig:greenCloudTaxVAT}
 \end{center}
\end{figure}

We executed twelve market scenarios for evaluating the GreenCloud tax. Selected simulation results are shown in figure~\ref{fig:scenarioResults}. The providers $P1-6$ run virtual machines for consumers in a market scenario where a value-added tax (VAT) with a tax rate of $10\%$ is used. The datacenters of the other providers are idle. This is because the value-added tax does not give consumers an incentive to host virtual machines at providers that run energy-efficient servers. Figure~\ref{fig:greenCloudTaxVAT} shows a schematic overview of the prices of a virtual machine using the value-added tax. Provider $P1$ offers the cheapest virtual machine, while $P15$ offers the virtual machine at the highest price. The virtual machines of the providers with inefficient services are taxed at $10\%$ as well as the virtual machines that are hosted by providers running energy-efficient servers. The prices of the providers running energy-efficient servers are higher than the prices of providers running energy-inefficient servers. So the value-added tax increases the price difference - consider the following example: a provider charges a price of $10\$$ for a virtual machine, while a provider with energy-efficient servers charges $15\$$ for the identical virtual machine, the difference of the prices is $\Delta=5\$$. With the value-added tax of $10\%$, the difference increases to $\Delta=5.5\$$.

\begin{figure}
 \begin{center}
    \includegraphics[width=0.6\linewidth]{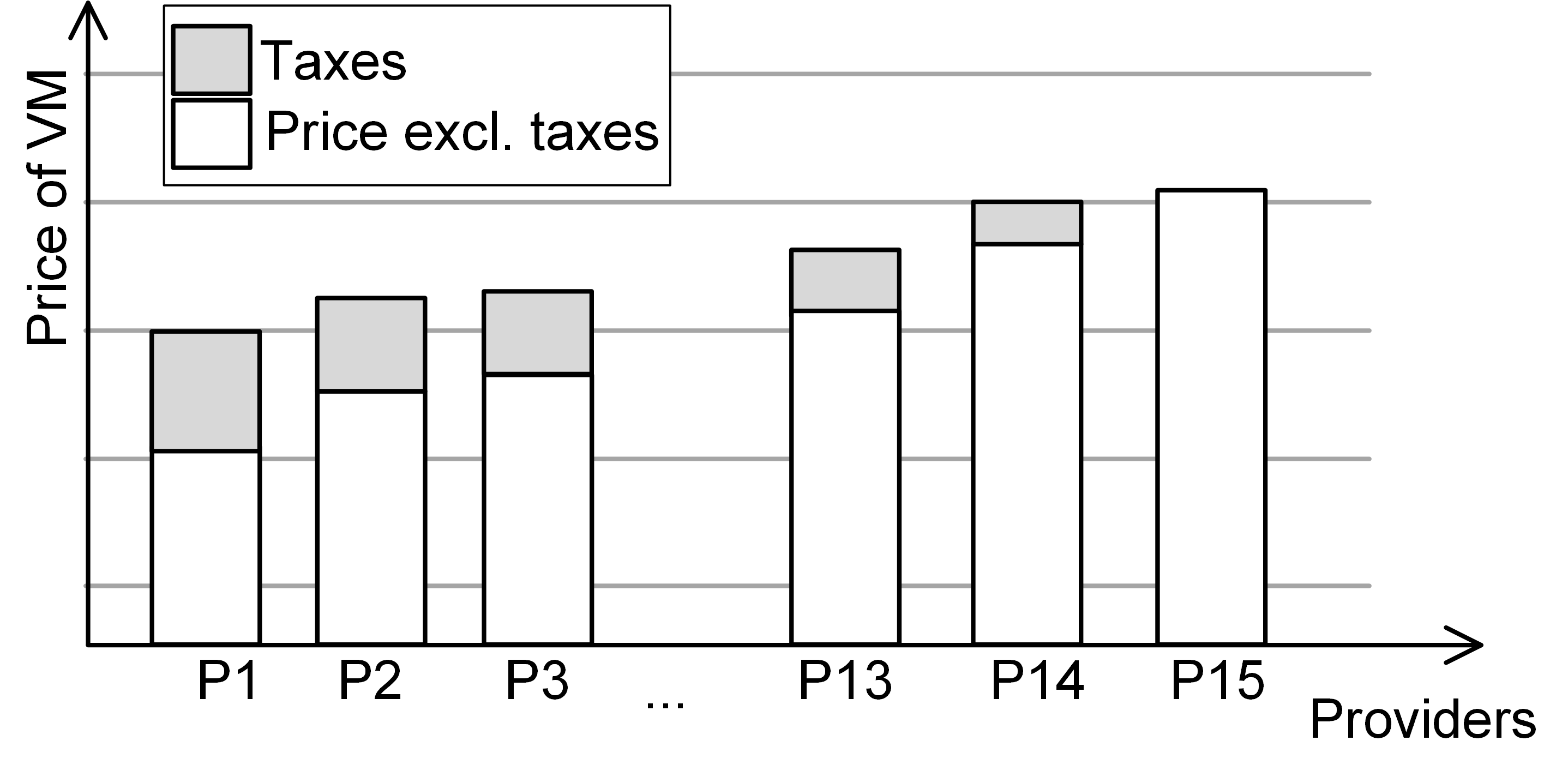}
    \caption{Schematic overview of the prices of a virtual machine using the GreenCloud tax - tax depends on the energy efficiency of the servers hosting the virtual machine and the price}
    \label{fig:greenCloudTax}
 \end{center}
\end{figure}

The GreenCloud tax was used for the other market scenarios that we executed.  Figure~\ref{fig:scenarioResults} shows that the GreenCloud tax with an eco penalty (EP) of $0.9$ has no significant effect on the market: the providers with the inefficient servers $P1-6$ still host all consumers even if they are higher taxed than the remaining providers. However, the eco penalty and consequently the tax size are too small to give consumers an incentive to use virtual machines from providers that run energy-efficient servers. A schematic overview of the prices of virtual machines using the GreenCloud tax is depicted in figure~\ref{fig:greenCloudTax}. Here, the highly taxed price of provider $P1$ is still cheaper than that of e.g., provider $P15$. An eco penalty of $1.09$ and $1.1$ leads to a tax size that gives consumers an incentive to move to providers that run more energy-efficient servers. In the market scenario where the GreenCloud tax with an eco penalty of $1.2$ was used, only the providers with the most energy-efficient servers $P10-15$ host the virtual machines. The virtual machines of the other providers are so highly taxed that no consumers buy them. A lot of consumers leave the market using the GreenCloud tax with an eco penalty of $80$. Here, only the providers $P13-15$ with the most energy-efficient servers host virtual machines.

%

%

\begin{figure*}
 \begin{center}
    \includegraphics[width=1\linewidth]{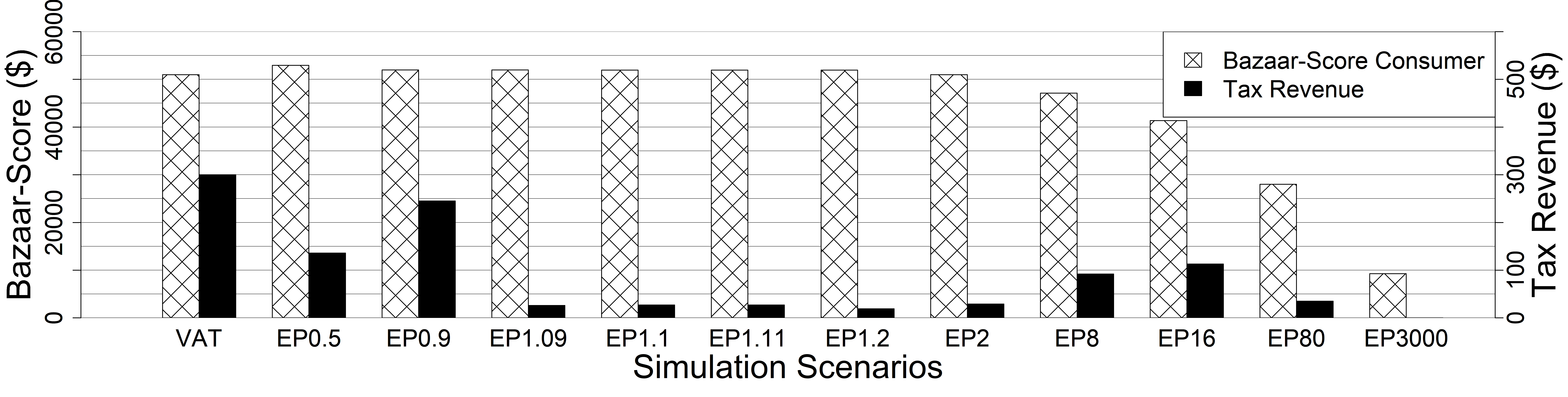}
    \caption{Simulation results - tax revenue and the Consumer Bazaar-Score (measuring the consumer welfare) of different tax systems, VAT=value-added tax, EP=eco penalty for the used GreenCloud tax}
    \label{fig:scenarioResultsTax}
 \end{center}
\end{figure*}

The scenarios show that the GreenCloud tax fosters providers with energy-efficient servers.  However, from the tax authorities' point of view, the GreenCloud tax comes at a price: The GreenCloud tax leads to a disproportionately increment of the low prices offered by providers with inefficient servers, as depicted in figure~\ref{fig:greenCloudTax}. The minimum price in the scenario where the value-added tax is used (see figure~\ref{fig:greenCloudTaxVAT}) is lower than in the scenario where a GreenCloud tax is used (see figure~\ref{fig:greenCloudTax}). Consumers use the gross price for making their decisions - it does not matter if the tax authority or the provider gets the price paid. Hence, using the GreenCloud tax, consumers with a low willingness to pay leave the market earlier than in a market where the value-added tax is used. In other words, the low prices increase faster when the GreenCloud tax is applied. This affects the tax revenue - defined in section~\ref{sec:greencloudtax} - as figure~\ref{fig:scenarioResultsTax} shows. It has a maximum using the value-added tax of $10\%$.  
A significant loss of the tax revenue can be observed by increasing the eco penalty from 0.9 to 1.09 in figure~\ref{fig:scenarioResultsTax}. Here, the tax revenue drops from $250\$$ to less than $50\$$. This neither implies that the consumers pay less taxes for the virtual machines nor that a lot of consumers left the market. 
Indeed, figure~\ref{fig:scenarioResults} shows that this increment of the eco penalty has a significant effect on the consumers' decision: while in the market scenario with an eco penalty of $0.9$, the consumers use the inefficient providers $P1-6$, most of the consumers move to the most efficient providers in the scenario with an eco penalty of $1.09$. The GreenCloud tax does not tax the efficient providers very high, and so the tax authority does not gain a high tax revenue if consumers use the most efficient providers. In other words, the consumers move to providers with energy-efficient servers, which have higher net prices, but which are taxed low. So at the end, the consumers do not pay significantly more than in the scenario with an eco penalty of $0.9$, which the Bazaar-Score\footnote{the Consumer Bazaar-Score is a key metric for measuring the consumers' welfare and was introduced in~\cite{PittlMS16}} shown in figure~\ref{fig:scenarioResultsTax} illustrates. 

Figure~\ref{fig:scenarioResultsTax} further shows that the tax revenue suddenly increases for the eco markups $2$, $8$, $16$. In these scenarios, the additional tax revenue gained by the consumers, which remains in the market, can compensate for the loss of tax that occurs because consumers leave the market. For the eco markups $2$, $8$, $16$, and $80$, the previously described Laffer Curve can be observed - see section~\ref{sec:greencloudtax}. The tax revenue starts not increasing before an eco penalty of $2$ because consumers can - as described before - move to more energy-efficient providers that are lower taxed. However, with an eco penalty equal to or greater than $2$, the consumers are already using the most energy-efficient providers. 

The Bazaar-Score of all GreenCloud tax scenarios shows that with an increasing eco penalty and consequently with an increasing tax, the consumers' welfare is decreased. The higher the total price consumers have to pay, the less consumers will buy the virtual machine. For the remaining consumers, the welfare decreases as the introduced tax decreases the difference between price paid and willingness to pay -  see section~\ref{sec:greencloudtax}.

\begin{table*}
\small
\caption{Servers that are used by the providers P1-P15, data taken from the SPEC benchmark~\cite{spec_server_nodate}}
\begin{center}
\begin{tabular}{llc}
\hline
\hline
\textbf{Prov.} &  \textbf{Server} & \textbf{ssj\_ops/watt} \\
\hline
P1 &  Hewlett-Packard Company, ProLiant DL385 G5, AMD Opteron 2356 & 498 \\
P2 &  Acer Incorporated, Altos R520, Intel Xeon E5450  & 731 \\
P3 & Plat'Home, Cloud Station E, Intel Xeon L5530 & 1903 \\
P4 &  Acer Incorporated, Acer AR585 F1, AMD Opteron 6276 & 2076 \\
P5 & Tyan Computer, B8228Y190X2-045V4H, AMD Opteron 4376 HE & 3293 \\
P6 & Fujitsu, PRIMERGY RX200 S7, Intel Xeon E5-2660  & 4722 \\
P7 & Fujitsu, PRIMERGY TX120 S3p, Intel Xeon E3-1265LV2  & 6109 \\
P8 & Huawei Technologies, Tecal CH121, Intel Xeon E5-2660 v2  & 6453 \\
P9 &  Supermirco, SYS-1028R-WC1RT, Intel Xeon E5-2699 v3 & 8004 \\
P10 &  Sugon, Sugon I620-G20, Intel Xeon E5-2699 v3 2.30 GHz  & 9939 \\
P11 &  Hewlett Packard, ProLiant DL360 9, Intel Xeon E5-2699 v3 2.30 GHz  & 10118 \\
P12 &  Sugon, I620-G30, Intel Xeon Platinum 8180 @2.50 GHz   & 11568 \\
P13 &  Hewlett Packard, ProLiant DL360 Gen10, Intel Xeon 8180 2.50GHz  & 12313 \\
P14 &  Lenovo Global, ThinkSystem SR650, Intel Xeon 8176 CPU 2.10 GHz    & 12336 \\
P15 &  H3C, UniServer R4900 G3, Intel(R) Xeon(R) 8180 CPU @2.50GHz & 12368 \\
\hline
\end{tabular}
\end{center}
\label{tab:servers}
\end{table*}

\subsection{Discussion}
The introduced market scenarios show the benefits and drawbacks of the GreenCloud tax. The governments face a trade-off between tax revenue and ecological efficiency. We also have to mention that the underlying market structure - a Bazaar-based market in our case - influences the results. For example, the negotiation strategy determines when to stop negotiation and which prices are accepted. Nevertheless, the basic trends analyzed in the previous section are independent of the underlying market structure. Further, it has to be considered that the GreenCloud tax can be realized in different ways. E.g., the most efficient provider - which was not taxed in the previous scenarios - can also be taxed, which helps to increase the tax revenue. A main benefit is that the GreenCloud tax does not consider the current costs of energy. Independent of the current energy prices, consumers have an incentive to purchase virtual machines from providers that run energy-efficient servers.

We see the development of widely accepted metrics for measuring the energy efficiency of servers as a main challenge towards a GreenCloud tax. Further, tax authorities have to accept a possible reduction of the tax revenue that occurs if consumers move to providers running energy-efficient servers.

\section{Conclusion and Further Research}
\label{sec:conclusion}

The energy demand of datacenters is significant. Latest forecasts imply significant improvements in energy efficiency. If these improvements can not be achieved, the energy demand will dramatically increase. Hence, in this paper, we introduced an economical instrument - an IaaS tax - for improving the energy efficiency of datacenters. Thereby, virtual machines that are hosted by providers that run energy-efficient servers face tax benefits, while virtual machines that are hosted by providers that run energy-inefficient servers are penalized. 
We evaluated the introduced tax model using a CloudSim-based simulation environment, which we developed. It allows for the simulation of Bazaar-based markets where different tax systems, such as the proposed tax model, can be evaluated. In the paper at hand, we used a real data set published in the SPEC benchmark for the simulations.

In our further research, we will further investigate the impact of such a tax, as well as additional metrics that can be used for evaluating the energy efficiency of servers. Moreover, we will analyze and develop more complex tax systems that consider the used type of computing resource.


\bibliography{greencloud}
\bibliographystyle{splncs03}

\end{document}